\documentclass[preprint2]{aastex}

\usepackage{natbib}
\usepackage{graphicx} 
\usepackage{subfigure}

\bibpunct{(}{)}{;}{a}{}{,}

\shorttitle{\Huge{MASSCLEAN}{\it colors} \\-- Mass Dependent Integrated Colors for Stellar Clusters --} 
\shortauthors{B. Popescu & M.M. Hanson}

\begin{document}

\slugcomment{DRAFT DATE: \today}

\title{\LARGE{MASSCLEAN}{\it colors} \\-- {\Large Mass Dependent Integrated Colors for Stellar Clusters Derived from 30 Million {\it Monte Carlo} Simulations} --}

\normalsize

\author{Bogdan Popescu and M. M. Hanson}
\affil{Department of Physics, University of Cincinnati, PO Box 210011, Cincinnati, OH 45221-0011}
\email{popescb@mail.uc.edu, margaret.hanson@uc.edu}

\begin{abstract}

We present Monte Carlo models of open stellar clusters with the purpose of mapping out the behavior of integrated colors with mass and age.  Our cluster simulation package allows for stochastic variations in the stellar mass function to evaluate variations in integrated cluster properties.  We find that $UBVK$ colors from our simulations are consistent with simple stellar population (SSP) models, provided the cluster mass is large, $M_{cluster} \ge 10^6 M_{\odot}$.  Below this mass, our simulations show two significant effects.  First, the {\sl mean value} of the distribution of integrated colors moves away from the SSP predictions and is less red, in the first 10$^7$ to 10$^8$ years in $UBV$ colors, and for all ages in $(V-K)$.  Second, the $1 \sigma$ dispersion of observed colors increases significantly with lower cluster mass.  The former we attribute to the reduced number of red luminous stars in most of the lower mass clusters and the later we attribute to the increased stochastic effect of a few of these stars on lower mass clusters.  This later point was always assumed to occur, but we now provide the first public code able to quantify this effect. We are completing a more extensive database of magnitudes and colors as a function of stellar cluster age and mass that will allow the determination of the correlation coefficients among different bands, and improve estimates of cluster age and mass from integrated photometry.






\vskip 0.5cm


{\small Submitted to {\it The Astrophysical Journal, Letters}}

\vskip 0.5cm

\end{abstract}


\section{Introduction}

When dealing with the integrated light from extremely massive systems, such as entire galaxies, one can reasonably assume there are enough stars in the system to represent all critical stages of evolution, regardless of stellar mass or age. When studying very massive stellar clusters, 100,000 $M_{\Sun}$ or more, while all members are (typically) of the same age, the isochrones will be fully populated at all ages (e.g. \citeauthor*{bruzual2003} \citeyear{bruzual2003}).  

However, typical open star clusters do not have such high masses.  Two open star clusters of nearly identical age can not be expected to have the same integrated light properties. This is because of natural cluster-to-cluster variations in the exact stellar masses found within the cluster, which give way to the cluster's integrated light properties (e.g. \citeauthor*{cervino2009} \citeyear{cervino2009}; \citeauthor*{buzzoni} \citeyear{buzzoni}; \citeauthor*{chiosi} \citeyear{chiosi}). In their final evolutionary states, stars become very luminous  and/or possess rather extreme red or blue colors.  This can cause observable deviations in integrated colors of the cluster as compared to the integrated colors computed in the high mass limit.  This effect is particularly great in young to middle-aged open clusters where their dying stars are the most luminous and have the most extreme and rapid color changes.  This poses a serious problem to astronomers because integrated colors are the primary way in which the age and mass of unresolved stellar clusters are derived (e.g. \citeauthor*{larsen2009} \citeyear{larsen2009}).

The fact that integrated colors are affected by the finite value of cluster mass and the fluctuations in the IMF has been known and studied by many authors (e.g. \citeauthor*{bruzual2001} \citeyear{bruzual2001}; \citeauthor*{bruzual2009} \citeyear{bruzual2009}; \citeauthor*{cervino2004} \citeyear{cervino2004}; \citeauthor*{cervino2006} \citeyear{cervino2006}; \citeauthor*{fagiolini2007} \citeyear{fagiolini2007}; \citeauthor*{lancon2000} \citeyear{lancon2000}; \citeauthor*{lancon2009} \citeyear{lancon2009}; \citeauthor*{fouesneau} \citeyear{fouesneau}; \citeauthor*{brocato2000a} \citeyear{brocato2000a}; \citeauthor*{brocato2000b} \citeyear{brocato2000b}; \citeauthor*{buzzoni} \citeyear{buzzoni}; \citeauthor*{chiosi} \citeyear{chiosi}; \citeauthor*{cantiello} \citeyear{cantiello}; \citeauthor*{raimondo2005} \citeyear{raimondo2005}; \citeauthor*{raimondo2009} \citeyear{raimondo2009}; \citeauthor*{piskunov} \citeyear{piskunov}; \citeauthor*{santos} \citeyear{santos}; \citeauthor*{gonzalez} \citeyear{gonzalez}; \citeauthor*{gonzalez2005} \citeyear{gonzalez2005}; \citeauthor*{gonzalez2010} \citeyear{gonzalez2010}; \citeauthor*{jesus} \citeyear{jesus}; \citeauthor*{pessev} \citeyear{pessev}). 
We present a large number of Monte Carlo simulations, which constitute the MASSCLEAN{\it colors} database, using the \texttt{MASSCLEAN (\textbf{MASS}ive \textbf{CL}uster \textbf{E}volution and \textbf{AN}alysis)} package\footnote{\url{http://www.physics.uc.edu/\textasciitilde popescu/massclean/}} (\citeauthor*{masscleanpaper} \citeyear{masscleanpaper}). This is a public and open source code. While it was first designed to simulate particular clusters, new additions to the package allow it to run in Monte Carlo mode. Also, all of the tools used to perform statistical analysis described in this work are included, so we are able to directly predict the mean value of the distribution of colors and color dispersion {\sl as a function of cluster mass} and age.


\texttt{MASSCLEAN} is a new stellar cluster image and photometry simulation package.  It uses an algorithm which populates the simulated cluster with a discrete number of tenable stars and then evolves each individual star following isochrones (Padova -- \citeauthor*{padova2008} \citeyear{padova2008} or Geneva -- \citeauthor*{geneva01} \citeyear{geneva01}; \citeauthor*{geneva02} \citeyear{geneva02} with isochrones obtained using Basel atmosphere libraries as presented in \citeauthor*{geneva1} \citeyear{geneva1}).  Integrated photometry of the simulated cluster is derived by simply summing up the stellar flux at any point in the cluster evolution.  We have tested that at very high cluster mass, the ``infinite mass limit'' of  M$_{cluster}$ $\ge$ 10$^6$ $M_{\Sun}$, the integrated colors are consistent with standard simple stellar population (SSP) models when using identical cluster properties of age, IMF,  evolutionary code, etc.  (\citeauthor*{masscleanpaper} \citeyear{masscleanpaper}).  In this {\it Letter}, we present results using {\texttt{MASSCLEAN} to explore the colors and color ranges of a stellar cluster as the mass drops below this limit, introducing {\sl mass as a variable} in this investigation.

\section{Results from 30 million Monte Carlo simulations}

We have computed the mean value of $(B-V)_{0}$, $(U-B)_{0}$ and $(V-K)_{0}$ integrated colors as a function of mass and age for a simple stellar cluster.  Our results -- MASSCLEAN{\it colors} database -- are based on over $30$ million Monte Carlo simulations\footnote{corresponding to over $200,000$  \texttt{MASSCLEAN} runs (for a complete description of a \texttt{MASSCLEAN} run see \citeauthor*{masscleanpaper} \citeyear{masscleanpaper})}. Broadly speaking, the simulations provide $UBVRIJHK$ magnitudes as a function of age and mass.  For this demonstration we present just $UBVK$, for two metallicities, $Z=0.019$ (solar) and $Z=0.008$ (Large Magellanic Cloud, LMC) and employ the two most commonly used stellar evolution models, Padova (\citeauthor*{padova2008} \citeyear{padova2008}) and Geneva (\citeauthor*{geneva1} \citeyear{geneva1}).  
The simulations were done using Kroupa IMF (\citeauthor*{Kroupa2002} \citeyear{Kroupa2002}) with $0.1$ $M_{\Sun}$ and $120$ $M_{\Sun}$ mass limits. The age range of $[6.6,9.5]$ in $log(age/yr)$ was choosen 
to accomodate both Padova and Geneva models. An average of $5000$ clusters were simulated for each mass and age. 
These results are presented in Figures 1 through 5.  
Comparison with observational data from LMC and Milky Way clusters is discussed in \ref{data}.


\begin{figure*}[hp] 
\centering
\subfigure[$(B-V)_{0}$]{\includegraphics[angle=270,width=11.5cm, bb=55 50 500 758]{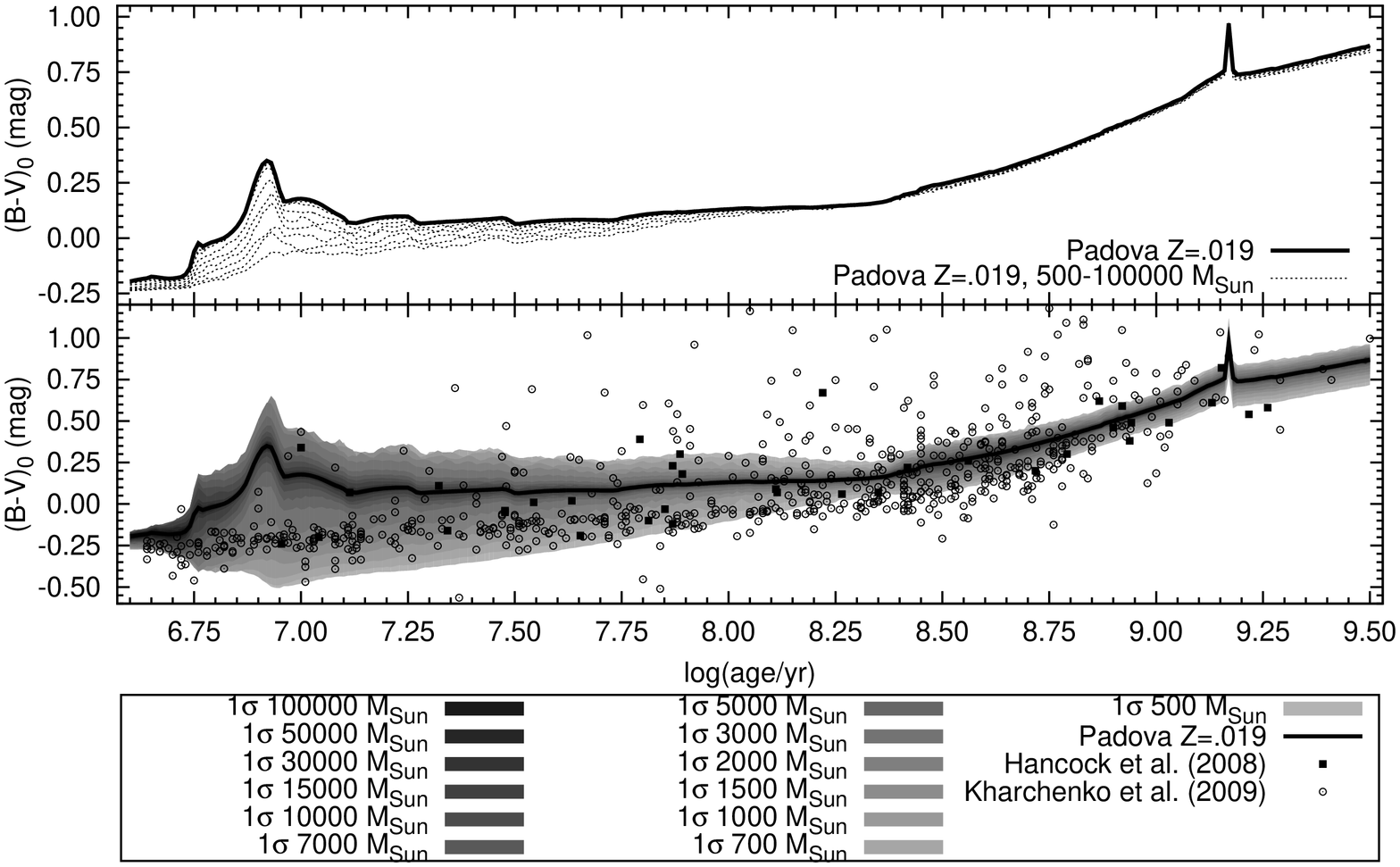}} 
\subfigure[$(U-B)_{0}$]{\includegraphics[angle=270,width=11.5cm, bb=55 50 408 758]{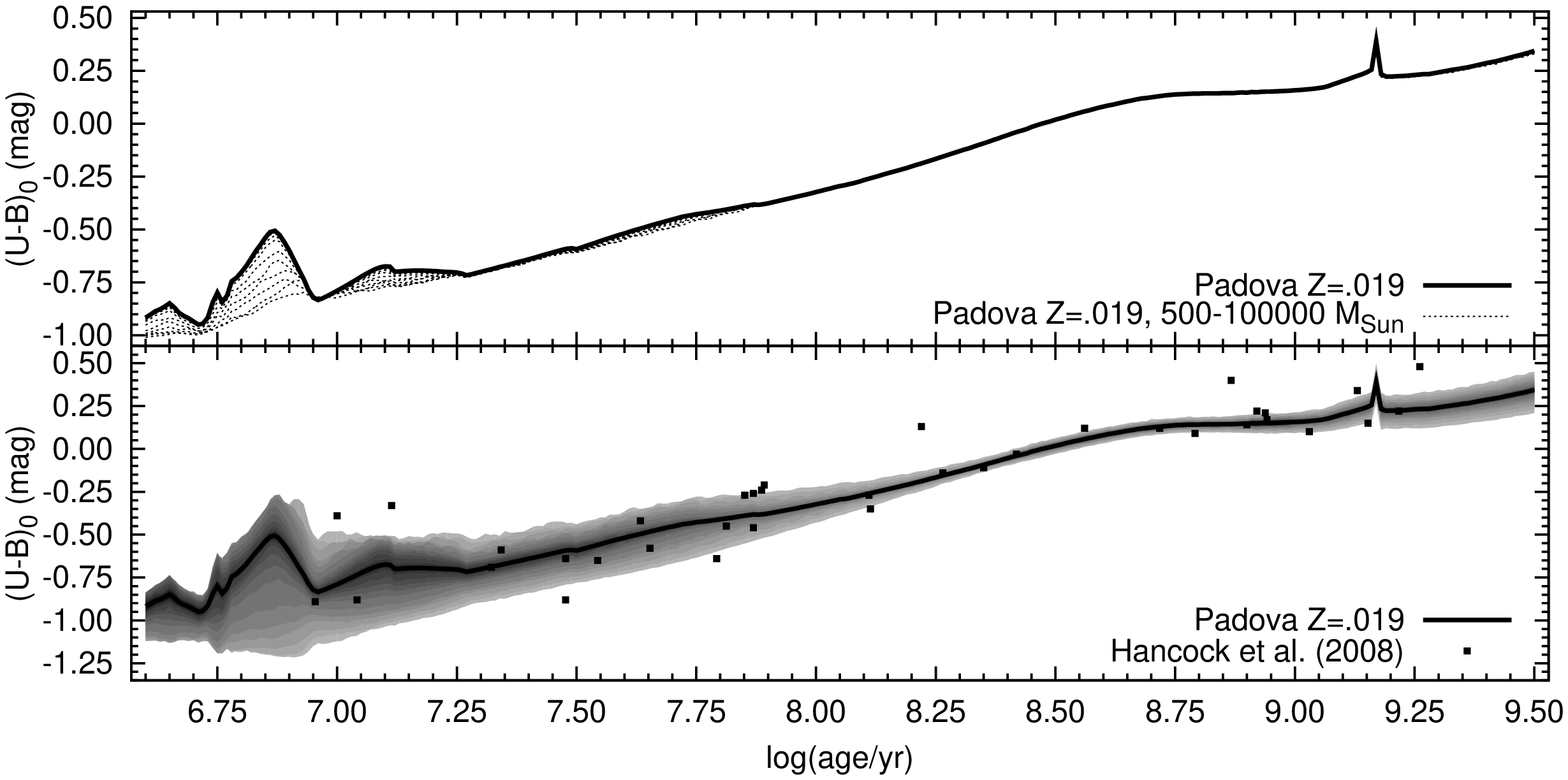}}
\subfigure[$(V-K)_{0}$]{\includegraphics[angle=270,width=11.5cm, bb=55 50 408 758]{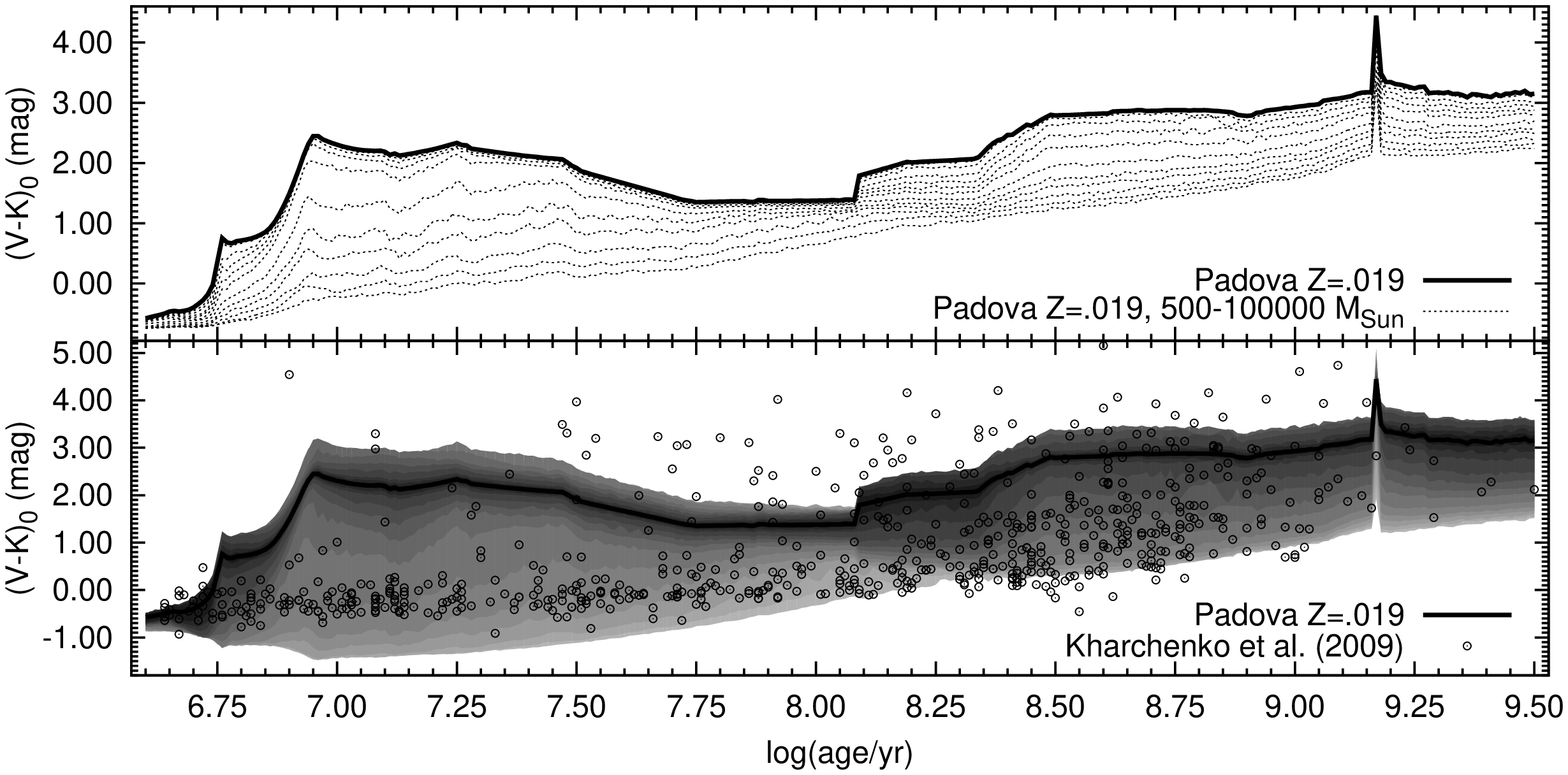}} 
\caption{\small \texttt{MASSCLEAN} integrated colors for Padova models $Z=0.19$ using $0.01$ $log(age/yr)$ step. The thick line corresponds to the infinite mass limit ($10^6$ $M_{\odot}$ in our simulations). {\it (a) Upper panel}: $(B-V)_{0}$ as a function of mass and age; {\it (b) Middle panel}: $(U-B)_{0}$ as a function of mass and age: {\it (c) Lower panel}: $(V-K)_0$ as a function of mass and age.   The Milky Way clusters are from \citeauthor*{hancock} \citeyearpar{hancock}  and \citeauthor*{kharchenko} \citeyearpar{kharchenko}.\normalsize}\label{fig:one}
\end{figure*}

\begin{figure*}[hp] 
\centering
\subfigure[$(B-V)_{0}$]{\includegraphics[angle=270,width=11.5cm, bb=55 50 500 758]{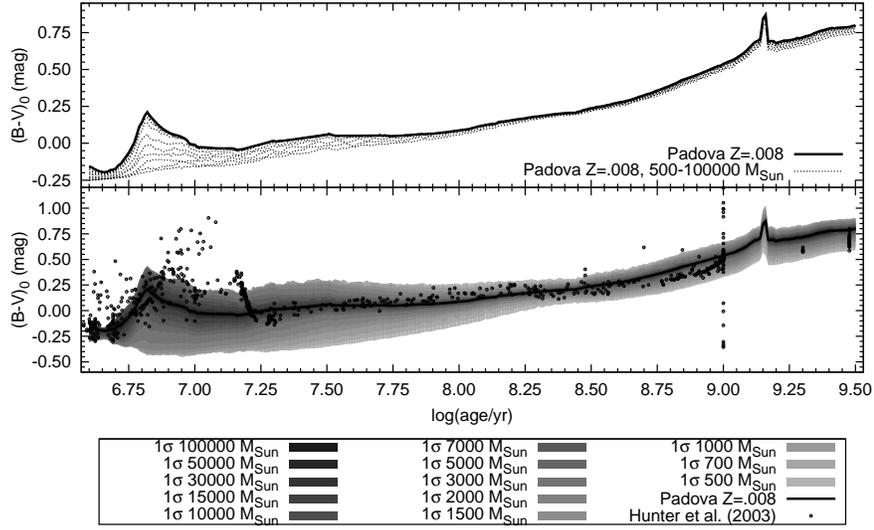}} 
\subfigure[$(U-B)_{0}$]{\includegraphics[angle=270,width=11.5cm, bb=55 50 408 758]{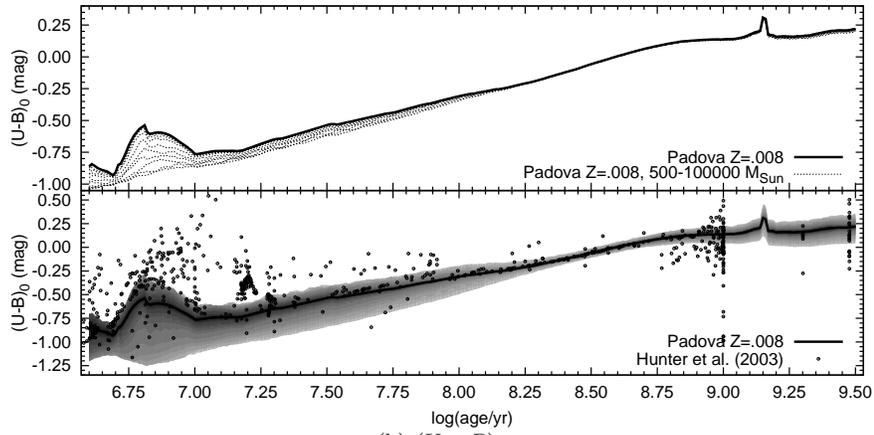}}
\subfigure[$(V-K)_{0}$]{\includegraphics[angle=270,width=11.5cm, bb=55 50 408 758]{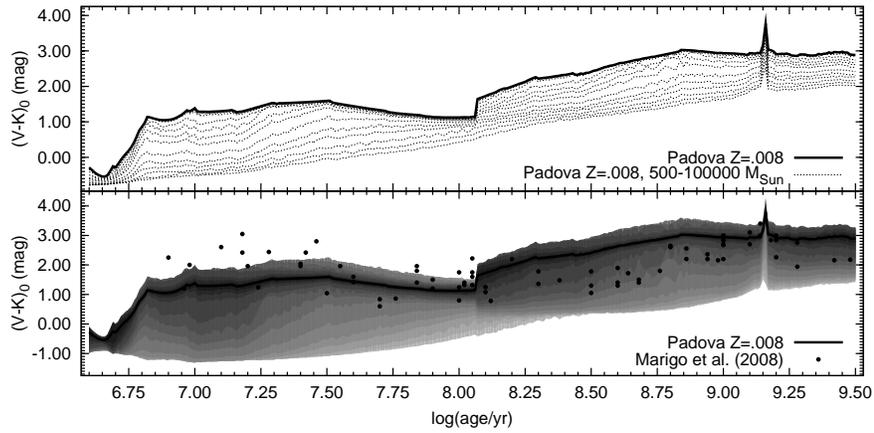}} 
\caption{\small \texttt{MASSCLEAN} integrated colors for Padova models $Z=0.008$ using $0.01$ $log(age/yr)$ step. The panels are the same as in Fig.\ 1, except for the data points.  Here, the clusters are from the LMC and found in  \citeauthor*{hunter2003} \citeyearpar{hunter2003} and Marigo et al.\ (2008). \normalsize}\label{fig:two}
\end{figure*}

\begin{figure*}[hp] 
\centering
\subfigure[$(B-V)_{0}$]{\includegraphics[angle=270,width=11.5cm, bb=55 50 500 758]{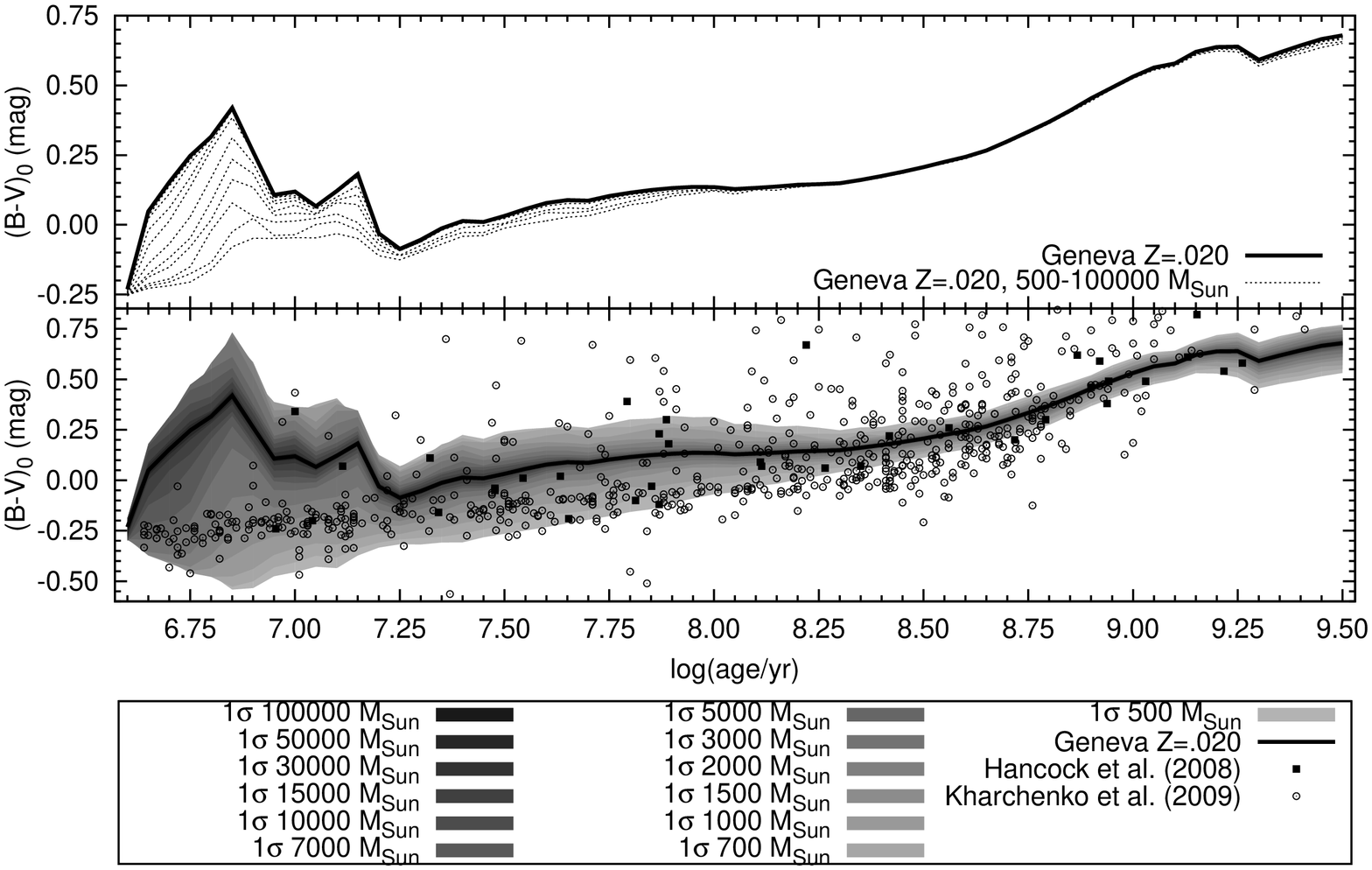}} 
\subfigure[$(U-B)_{0}$]{\includegraphics[angle=270,width=11.5cm, bb=55 50 408 758]{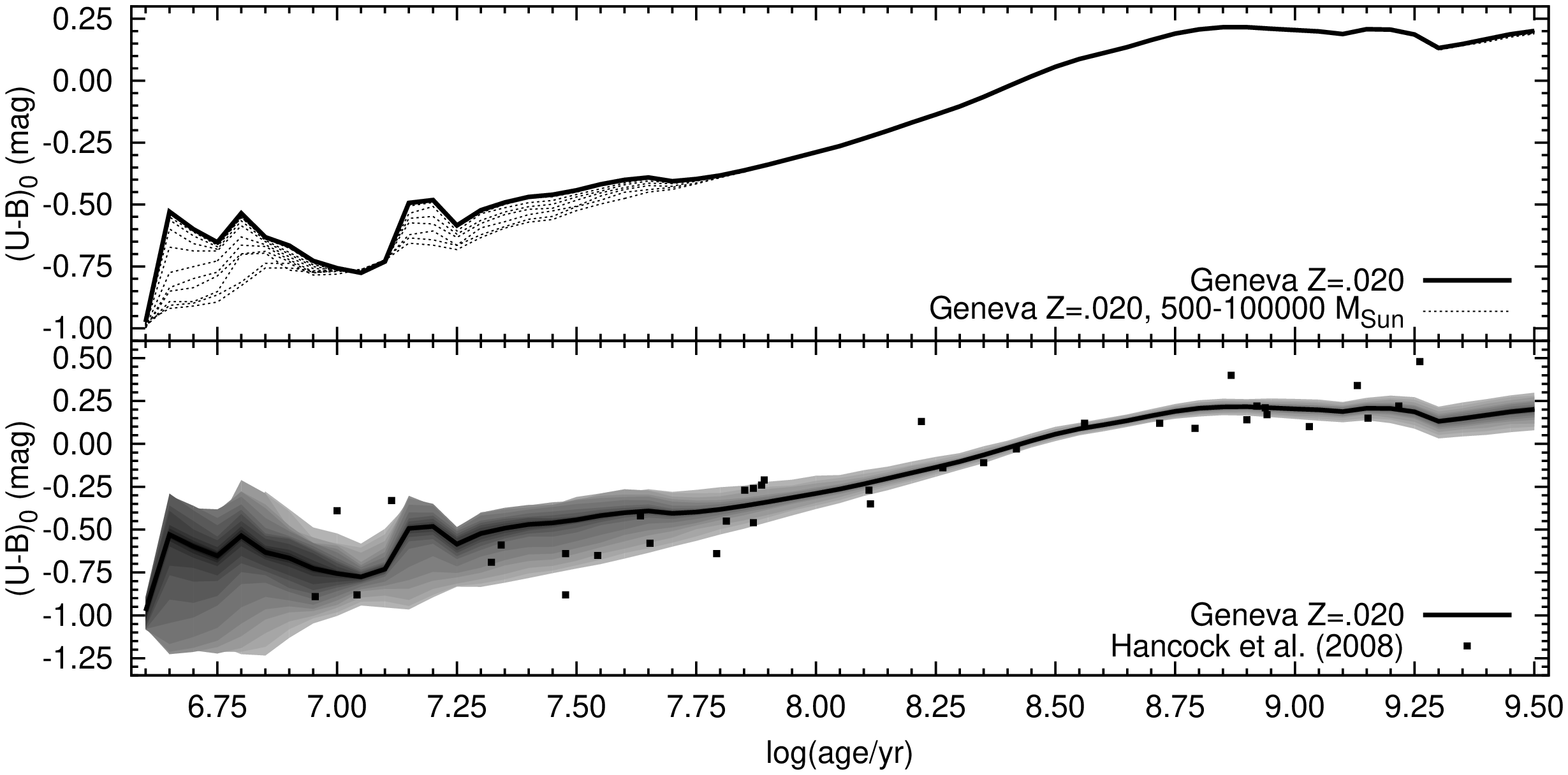}}
\subfigure[$(V-K)_{0}$]{\includegraphics[angle=270,width=11.5cm, bb=55 50 408 758]{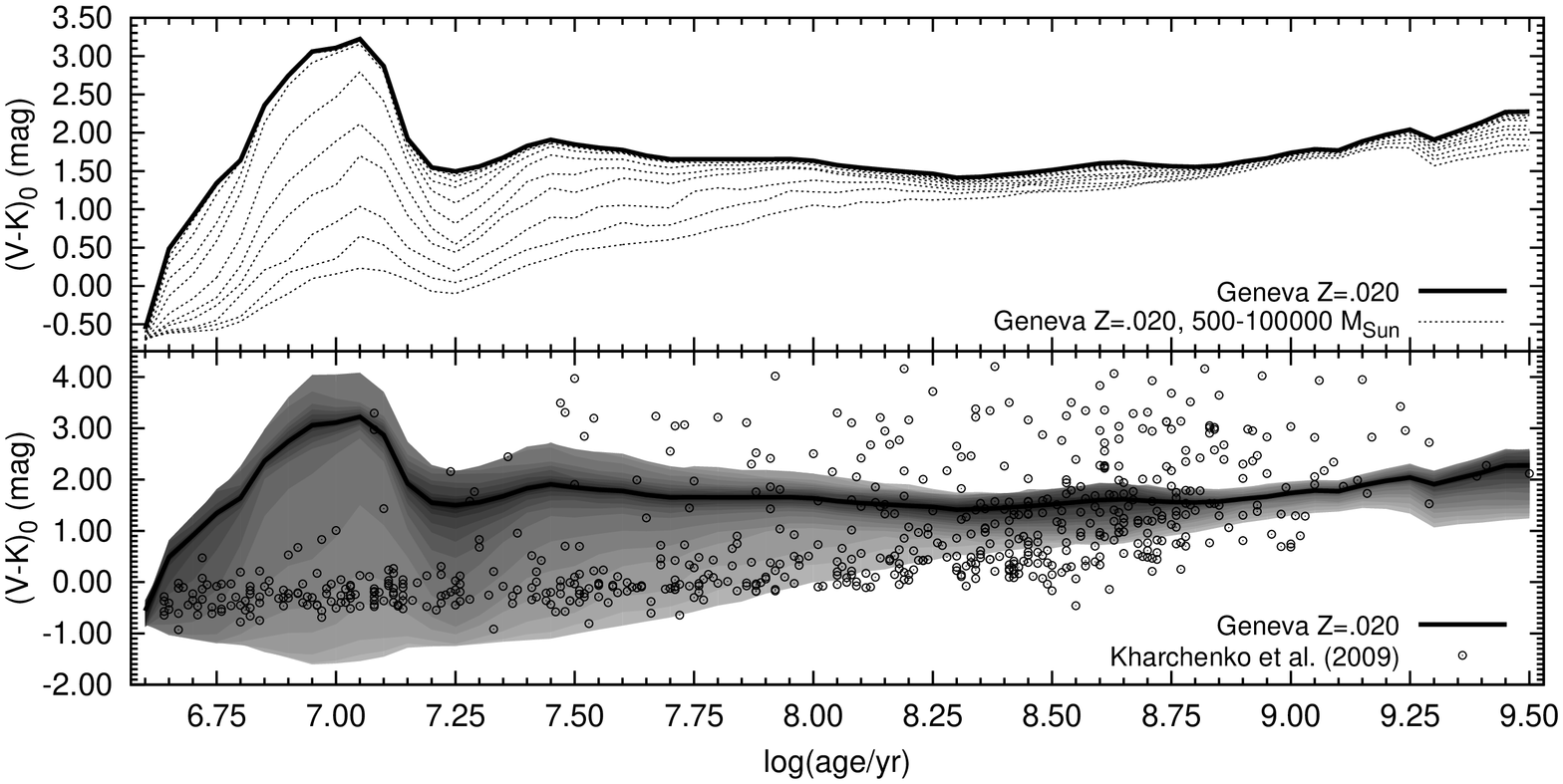}} 
\caption{\small \texttt{MASSCLEAN} integrated colors for Geneva models $Z=0.20$ using $0.05$ $log(age/yr)$ step.  The panels are the same as in Fig.\ 1, including the Milky Way clusters shown. \normalsize}\label{fig:three}
\end{figure*}

\begin{figure*}[hp] 
\centering
\subfigure[$(B-V)_{0}$]{\includegraphics[angle=270,width=11.5cm, bb=55 50 500 758]{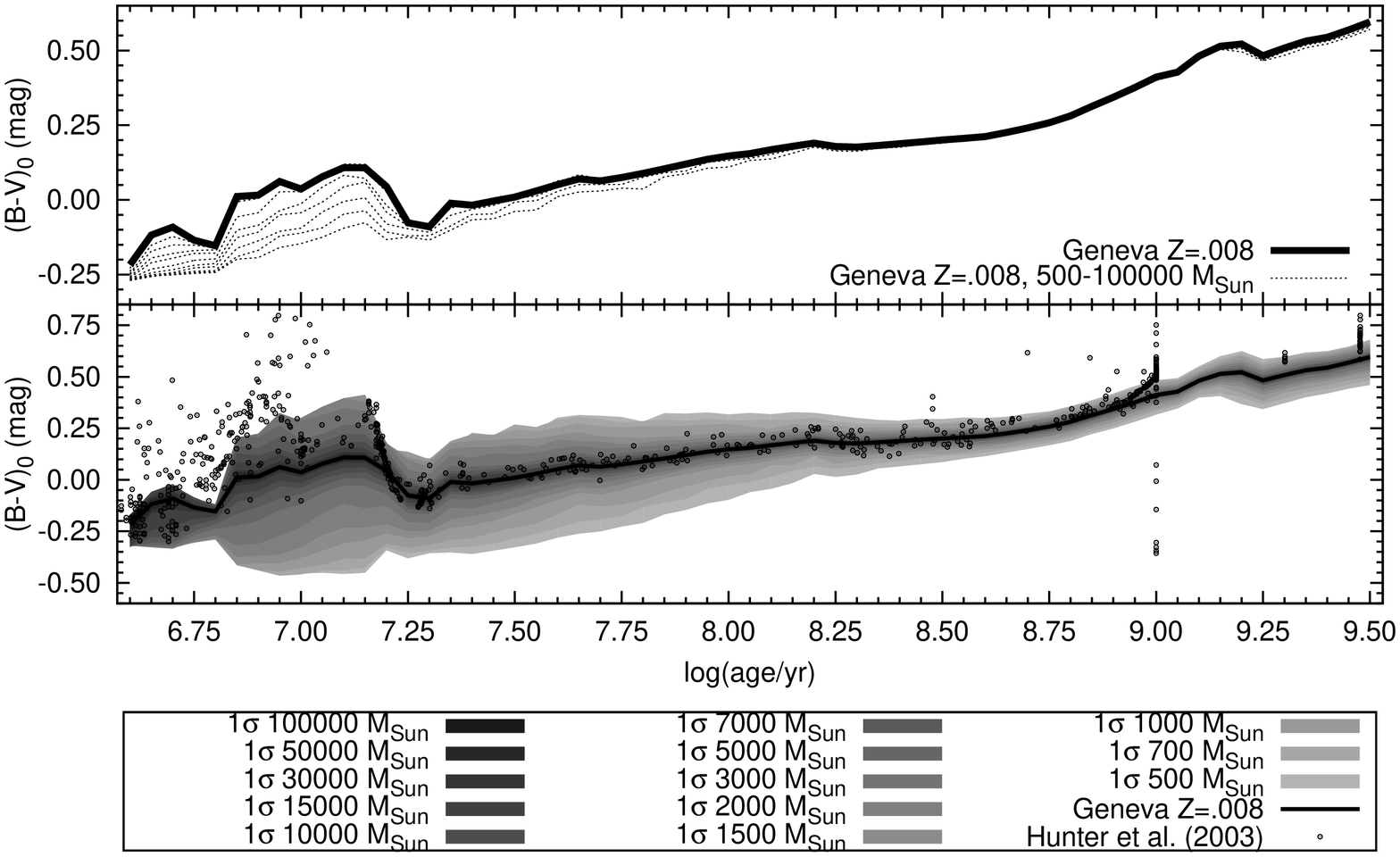}} 
\subfigure[$(U-B)_{0}$]{\includegraphics[angle=270,width=11.5cm, bb=55 50 408 758]{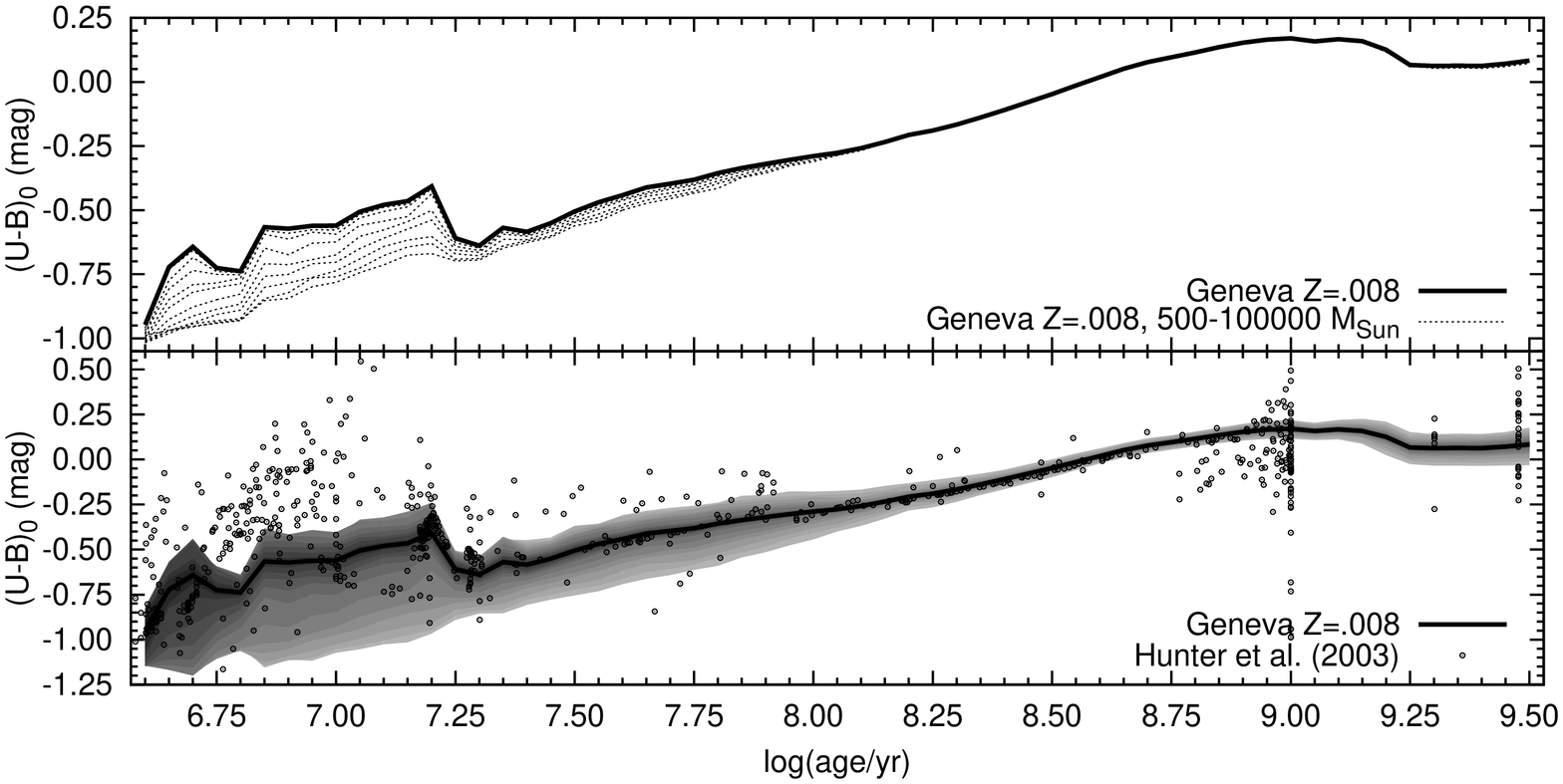}}
\subfigure[$(V-K)_{0}$]{\includegraphics[angle=270,width=11.5cm, bb=55 50 408 758]{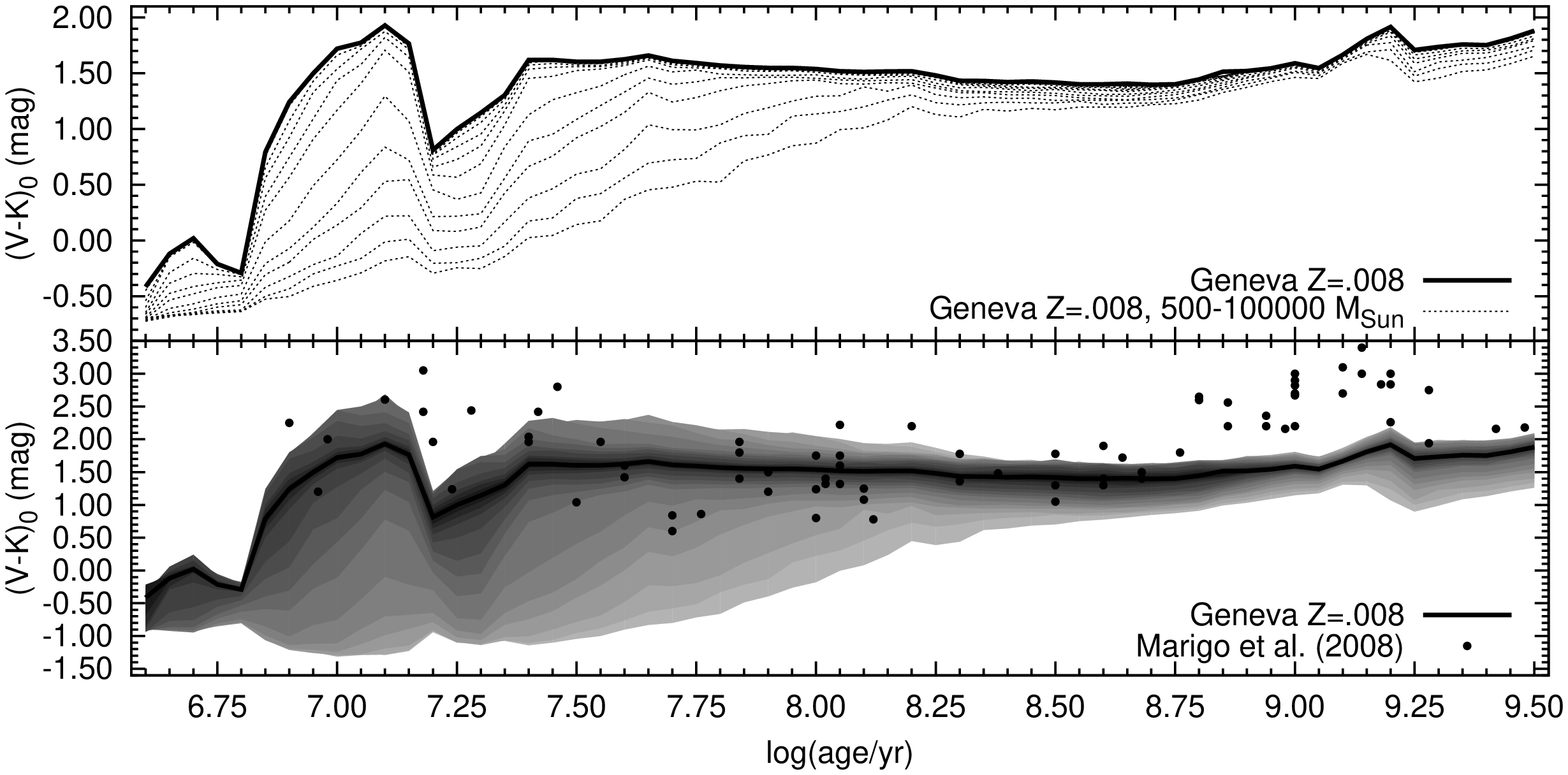}} 
\caption{\small \texttt{MASSCLEAN} integrated colors for Geneva models $Z=0.008$ using $0.05$ $log(age/yr)$ step. The panels are the same as in Fig.\ 1, except for the data points which are from the LMC. \normalsize}\label{fig:four}
\end{figure*}

\subsection{The ``blueing'' of mean integrated colors for lower mass clusters}
Figures \ref{fig:one} through \ref{fig:four}, are all similarly presented.  The {\bf mean value} for $(B-V)_{0}$, $(U-B)_{0}$ and $(V-K)_{0}$ distribution of integrated colors as generated by \texttt{MASSCLEAN} are given in the top of each of the three panels in each figure. In Figures \ref{fig:one} and \ref{fig:two},  we use the Padova evolutionary models with $log(age/yr)$ step of $0.01$, solar and LMC metallicity, respectively. In Figures \ref{fig:three} and \ref{fig:four}, we present the mean values using the Geneva evolutionary models with $log(age/yr)$ step of $0.05$, solar and LMC metallicity, respectively.  
The difference in smoothness seen in the Padova and Geneva sets is due to a smaller $log(age/yr)$ step in the Padova case ($0.01$, versus $0.05$ for Geneva).

The upper graph in each of the 3 panels (a) of Fig. \ref{fig:one} through \ref{fig:four} show the mean value of the integrated cluster color as a function of age.  Thirteen masses are presented starting at 10$^5$ $M_{\odot}$ (dark solid line, and virtually the same as the SSP codes predictions) and going down to 500 $M_{\odot}$ (dotted lines).  The dotted lines show a trend away from the SSP predicted values to bluer mean colors, as the cluster mass decreases for young clusters up to an age of as much as 10$^8$ years.   This occurs with both solar and LMC metallicities and both evolutionary models, Geneva and Padova.  More striking, as the mass decreases our simulations show a blueing of the integrated $(V-K)_0$ color  for all cluster ages simulated (bottom panel). 

\subsection{The increased range in the observed integrated colors of low mass clusters}
 The color dispersion (presented as $1 \sigma$ -- standard deviation -- about the mean) for different values of mass in the $500-100,000$ $M_{\Sun}$ interval is presented in the lower graphs of each panel.  Here we use differing levels of grayscale tone to represent the cluster masses.  The  most massive clusters show their $1 \sigma$ range in the darkest grayscale, while lower mass clusters are shown with increasingly lighter shades.  Recall, the mean value of the {\it distribution of integrated colors} moves to bluer colors as the cluster mass decreases.  This explains in part the greater color range seen in the low mass clusters on the blue side.  
 
 The expected number of stars in any given evolutionary
phase is fractional. This fact is not relevant if the cluster has
enough stars (is massive enough) so variations of $+/-$ one in a given
evolutionary phase have no impact. However, when the mass of the cluster 
decreases, this number becomes lower than one in the red, luminous
evolutionary phases, so most of the clusters in the simulation will have
no luminous red stars at all, since the number of stars in a given phase
must be an integer. This produces a bluer mean color. However, a few
clusters will have one (or a few) luminous red stars, so the result will be an excess in both sides (blue and red) (\citeauthor*{lancon2002} \citeyear{lancon2002}; \citeauthor*{cervino2004} \citeyear{cervino2004}; \citeauthor*{fagiolini2007} \citeyear{fagiolini2007}). 
In this situation it is difficult to obtain
a clear correlation of a particular (real) cluster integrated color with mass for a single age (\citeauthor*{santos} \citeyear{santos}; \citeauthor*{cervino2004} \citeyear{cervino2004}; \citeauthor*{cervino2006} \citeyear{cervino2006}; \citeauthor*{fagiolini2007} \citeyear{fagiolini2007}). While the correlation between the mean value (of the distribution of integrated colors) and mass is apparent in our plots, an accurate quantitative description could be obtained only using a larger number of simulations. However, we hope the predicted increased color range with lower mass should be a strong enough signal with a realistically observable number of clusters. In order to verify this results with observations, perhaps 100s of clusters would be needed and they must have {\sl well constrained} mass and age.

\subsection{The effect of low cluster mass on the UBV color-color diagrams}

The color-age diagrams  of Fig. \ref{fig:one} through \ref{fig:four} are not typically used by researchers because age is often not known for unresolved stellar clusters.  In fact, the great utility of SSP models has been to provide a way to derive cluster ages based on the easily observed integrated magnitudes and their location in a color-color diagram (\citeauthor*{searle} \citeyear{searle};\citeauthor*{girardi} \citeyear{girardi}). In Figure \ref{fig:five} we provide such diagrams.  Here we have re-plotted the simulations already shown in Fig. \ref{fig:one} through \ref{fig:four}, including the mean color lines and the greyscale $1 \sigma$ ranges of those colors, just like our earlier figures. Here the four panels represent the two evolutionary models (Padova and Geneva), with each of those using LMC and Milky Way metallicities.   We include the same real data shown in Figures \ref{fig:one} through \ref{fig:four} for the LMC and Milky Way.  Again we provide both mean colors as a function of age, and the $1 \sigma$ range of that mean color.  A similar color scheme for the range is employed, the darkest grayscale representing the most massive cluster simulated, while lighter shades represent lower mass clusters.  It is easy to see the very large dispersion of the lower mass clusters, particularly in the blue, young portion of Fig. \ref{fig:five}.  While its rather crowded, we've attempted to demonstrate the location of the mean color as a function of mass through the use of differing line styles in the plot.  As one might expect from our four earlier figures, most of the predicted mean variation occurs in the upper left portion of the diagram, corresponding to the youngest clusters.

\begin{figure*}[hp] 
\centering
\subfigure[Padova]{\includegraphics[angle=0,width=0.49\textwidth]{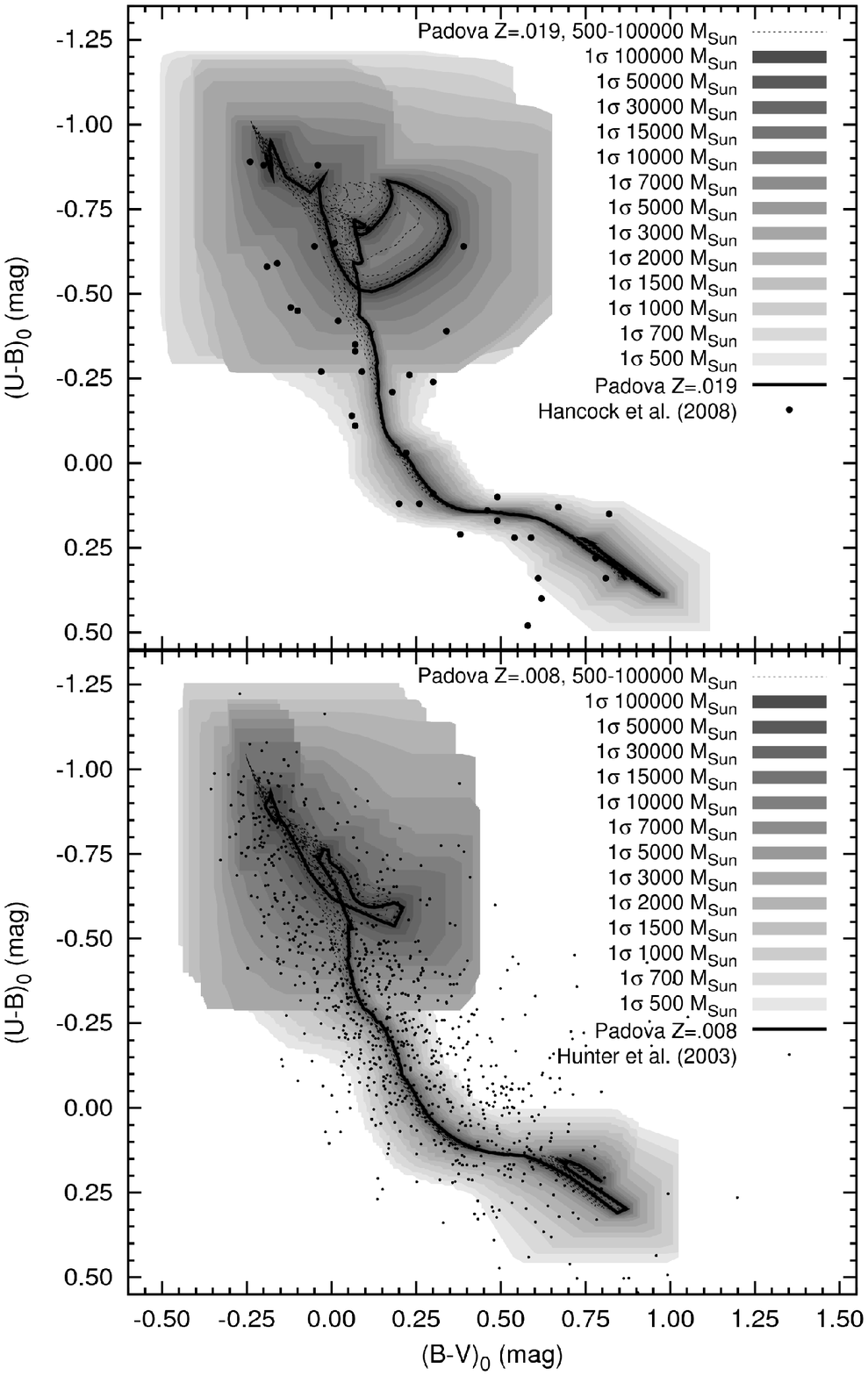}} 
\subfigure[Geneva]{\includegraphics[angle=0,width=0.49\textwidth]{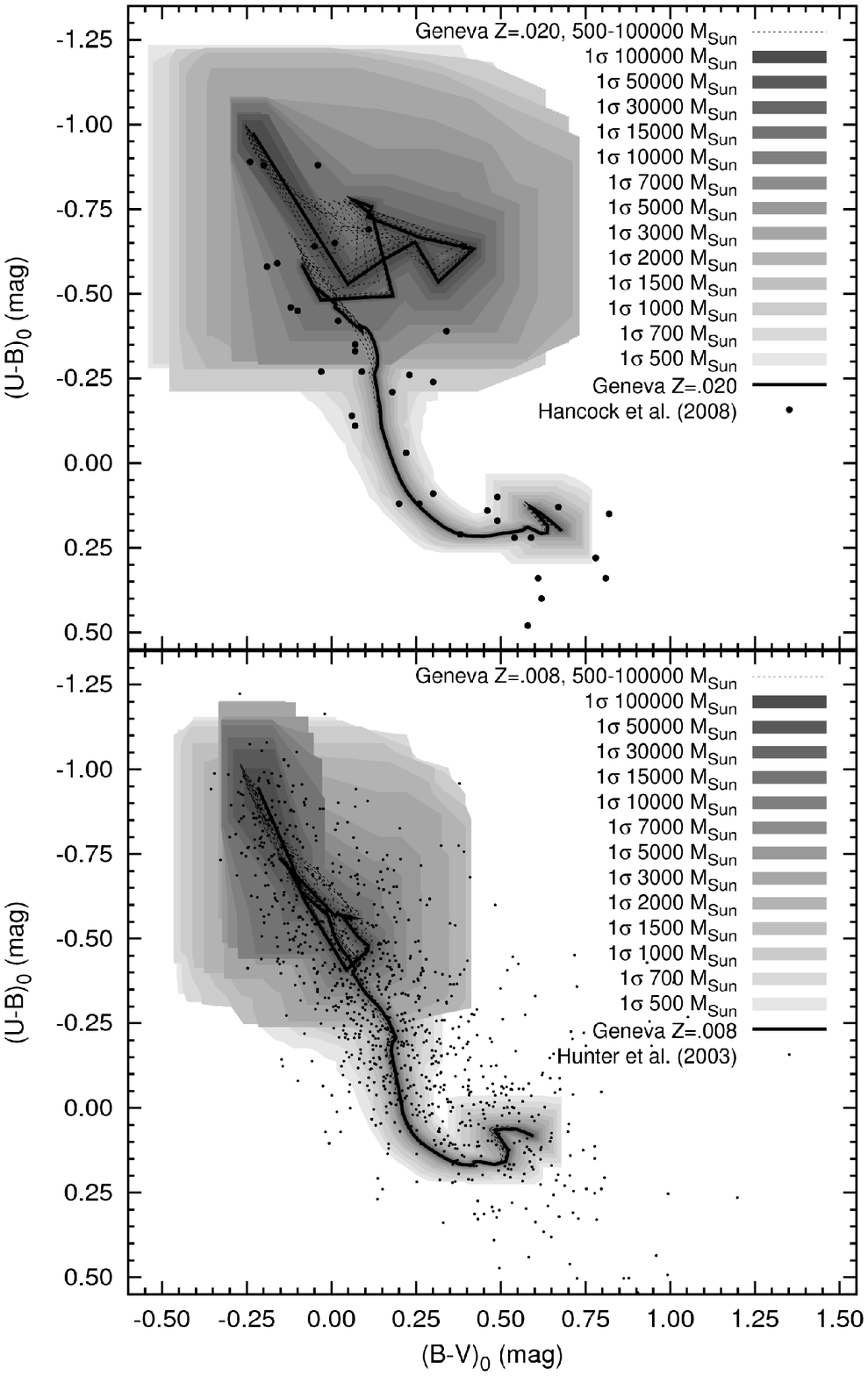}}
\caption{\small \texttt{MASSCLEAN} integrated $UBV$ color-color diagrams.  {\it (a)} Using Padova models $Z=0.19$ and $Z=0.008$ and {\it (b)} using Geneva models, $Z=0.20$ and $Z=0.008$.  The thick line corresponds to the infinite mass limit ($10^6$ $M_{\odot}$ in our simulations).  On top is given Milky Way clusters from \citeauthor*{hancock} \citeyearpar{hancock}  and the bottom panels provide data for LMC clusters, taken from 
\citeauthor*{hunter2003} \citeyearpar{hunter2003}.  \normalsize}\label{fig:five}
\end{figure*}

\subsection{Comparison with real cluster data} \label{data}
In Fig. \ref{fig:one} through \ref{fig:four}, we provide data taken from \citeauthor*{padova2008} \citeyearpar{padova2008} and \citeauthor*{hunter2003} \citeyearpar{hunter2003}, for the LMC  and  \citeauthor*{kharchenko} \citeyearpar{kharchenko}, and  \citeauthor*{hancock} \citeyearpar{hancock} for the Milky Way.  In Fig. \ref{fig:five}, we provide data from \citeauthor*{hunter2003} \citeyearpar{hunter2003}, for the LMC and \citeauthor*{hancock} \citeyearpar{hancock} for the Milky Way only.  The other datasets did not include $U$-band.  In Fig. \ref{fig:five}, there appears to be as large a variation of observed colors along the portion of the color-color diagram for the LMC clusters where old clusters are expected to be found (in the lower right) as young clusters (in the upper left). This seems incongruous with the rather tight fit of panels (a) and (b) in Figures \ref{fig:two} and \ref{fig:four}.  Also, in Fig. \ref{fig:five}, we see that the youngest clusters from the LMC are found to lie almost entirely within the gray scale error range when they do not do so in Fig. \ref{fig:one} through \ref{fig:four}.  The first is due to a selection effect of the data.  Clusters deviating greatly from the expected SSP curve in the $UBV$ color-color diagram of Fig. \ref{fig:five}, are not typically aged and thus will not appear in Fig. \ref{fig:two} and \ref{fig:four}, cleaning up those diagrams considerably. To understand why the younger clusters lie almost entirely within the grayscale of Fig. \ref{fig:five} and not Fig. \ref{fig:one} through \ref{fig:four}, one needs to recognize that in the blue range, the lines representing the mean double back on themselves, making the region very complex.  A quick check shows that the greatest range of grayscale in Fig. \ref{fig:five} never extends beyond the greatest range seen in Fig. \ref{fig:one} through \ref{fig:four}.

Looking at the figures one notes the data does not appear to be distributed in a manner consistent with our $1 \sigma$ ranges. In fact, one could say, the colors of real clusters shown often extend to values much redder than our $1 \sigma$ range given in the figures (more often than $1 \sigma$ would indicate for the data).  This is because of the way in which we have shown the color range with mass.  To provide a $1 \sigma$ range is most useful when the distribution is Gaussian.  Unfortunately, the distribution of observed colors is far from a well behaved Gauss about the mean for early ages and low mass clusters.  The distribution of colors with age in Fig. \ref{fig:one} through \ref{fig:four} becoming increasingly bimodal as mass decreases. This is consistent with the fluctuations described by \citeauthor*{lancon2000} \citeyearpar{lancon2000}; \citeauthor*{cervino2003} \citeyearpar{cervino2003}; \citeauthor*{cervino2004} \citeyearpar{cervino2004}; \citeauthor*{cervino2006} \citeyearpar{cervino2006}.  We will show that our simulations are consistent with these very red colors observed in low mass systems (Popescu \& Hanson 2010, in prep.).  We are running additional simulations and eventually will present more than 50 million clusters to better constrain the exact behavior of the colors as a function of mass and age.  

\section{Discussion and Conclusions}
While it is entirely expected that integrated colors computed by the modern SSP codes in the infinite mass limit will work best when applied to very massive systems, its been difficult to ascertain the error involved when one applies their predictions to lower mass systems. Our new stellar cluster simulation routine, \texttt{MASSCLEAN}, provides a means to determine this through Monte Carlo simulations of realistically modeled stellar clusters.  

We used \texttt{MASSCLEAN} to simulate 30 million stellar clusters over a mass range typical of open stellar clusters and for ages in the $[6.6, 9.5]$ $log(age/yr)$ range.  Our results indicate that as cluster mass decreases, the {\sl mean value of the distribution of colors} in the first tens (to hundreds) of millions of years will not reach to such high red values as predicted by SSP models.  Most of this very red color is brought on by very luminous red stars. Lower mass clusters will have fewer highly luminous red stars to pull the colors so red.  However, the stochastic variation of stars in these very luminous, quickly evolving phases, and thus very red colors, makes it still {\sl possible} just not so likely.  Thus we also find the typical range of colors observed increases significantly as the cluster mass decreases. This is because the color depends so strongly on the most luminous stars and a single star has the potential to produce greater deviations in the color of smaller clusters. Our results indicate a bluer mean value of the distribution of colors for lower mass clusters compared to SSP predictions in the infinite mass limit.

We continue to simulate still more clusters and expand the MASSCLEAN{\it colors} database to provide a complete sampling of colors and absolute magnitudes as a function of cluster age and mass (the work presented in this {\it Letter} was done during a six month period, using multiple computers). When complete, this database will provide a probabilistic grid from which a statistical inference code will be based to aid researchers in deriving cluster characteristics from their observed integrated photometry.  
We have completed some preliminary work to provide proof of concept (\citeauthor*{iaus266} \citeyear{iaus266}).   
With additional simulations (Popescu \& Hanson 2010, in prep.), we will be able to obtain the correlation coefficients among different bands.

\acknowledgements
We are grateful to suggestions made to an early draft of this work by Rupali Chandar and Deidre Hunter. 
We thank the referee for useful comments and suggestions.
This material is based upon work supported by the National Science Foundation under Grant No. 0607497 to the University of Cincinnati. 


\end{document}